\documentclass[11pt,a4paper]{article}
\usepackage[utf8]{inputenc}
\def\papertitle{\bf Probing direct $CP$ violation in $\Lambda_b^0 \to P_c^+ h^-$ $(h=\pi,K)$ with final-state rescattering} 
\usepackage{authblk}

\title{\papertitle}
\author[1]{Zhu-Ding Duan,\footnote{Email: duanzd@mail.imu.edu.cn}}
\author[2]{\,Tian-Liang Feng,\footnote{Email: fengtl18@lzu.edu.cn}}
\author[1]{\,Run-Hui Li,\footnote{Email: lirh@imu.edu.cn, corresponding author}}
\author[2]{\,Ming-Zhu Liu,\footnote{Email: liumz@lzu.edu.cn, corresponding author}}
\author[2,4]{\,Jian-Peng Wang,\footnote{Email: wangjp20@lzu.edu.cn, corresponding author}}
\author[2,3]{and Fu-Sheng Yu\footnote{Email: yufsh@lzu.edu.cn, corresponding author}}

\affil[1]{School of Physical Science and Technology, Inner Mongolia Key Laboratory of Microscale Physics and Atomic Manufacturing, Inner Mongolia University, Hohhot 010021, China}
\affil[2]{Frontiers Science Center for Rare Isotopes, and School of Nuclear Science and Technology, Lanzhou University, Lanzhou 730000, China}

\affil[3]{Lanzhou Center for Theoretical Physics, Key Laboratory of Theoretical Physics of Gansu Province, Key Laboratory of Quantum Theory and Applications of MoE, Gansu Provincial Research Center for Basic Disciplines of Quantum Physics, Lanzhou University, Lanzhou 730000, China}
\affil[4]{Physik Department, Universität Siegen, Walter-Flex-Str.3, D-57068 Siegen, Germany}
\usepackage{changes}
\usepackage[top=1in, bottom=1.25in, left=1in, right=1in]{geometry}
\usepackage{microtype}
\usepackage{lineno}
\usepackage{xspace}
\usepackage{booktabs}
\usepackage{comment}
\usepackage{slashed}

\usepackage{caption}

\usepackage{graphicx}  
\usepackage{color}
\usepackage{colortbl}

\DeclareGraphicsExtensions{.pdf,.PDF,png,.PNG}

\usepackage{amsmath} 
\usepackage{amsfonts,amssymb,amsthm,bm,braket,yhmath}
\usepackage{upgreek} 
\usepackage{hyperref}
\usepackage{hyperxmp}
\usepackage[figuresright]{rotating}
\usepackage{hypcap} 
\hypersetup{hidelinks} 
\usepackage{txfonts}
\usepackage{amssymb}
\usepackage{cases}
\usepackage{algorithm}
\usepackage{algpseudocode}
\usepackage{mathrsfs}
\usepackage{color}
\usepackage{amscd,amssymb,amsthm,amsmath,bm,graphicx,psfrag}
\usepackage{epsfig,mathrsfs}
\usepackage[bf,SL,BF]{subfigure}
\usepackage{amsmath}
\usepackage{pict2e,keyval,fp}

\theoremstyle{definition}

\numberwithin{equation}{section}

\def\kaon   {{\ensuremath{K}}\xspace}
\def\Kbar    {{\kern 0.2em\overline{\kern -0.2em \kaon}{}}\xspace}

\def\KorKbar    {\kern 0.18em\optbar{\kern -0.18em K}{}\xspace}

\def\Dbar    {{\kern 0.2em\overline{\kern -0.2em D}{}}\xspace}

\def\DorDbar    {\kern 0.18em\optbar{\kern -0.18em D}{}\xspace}

\def\Lbar        {{\ensuremath{\kern 0.1em\overline{\kern -0.1em\Lambda}}}\xspace}
\def\LorLbar    {\kern 0.18em\optbar{\kern -0.18em \PLambda}{}\xspace}

\newcommand{\tev}{\ensuremath{\mathrm{\,Te\kern -0.1em V}}\xspace}
\newcommand{\gev}{\ensuremath{\mathrm{\,Ge\kern -0.1em V}}\xspace}
\newcommand{\mev}{\ensuremath{\mathrm{\,Me\kern -0.1em V}}\xspace}
\newcommand{\kev}{\ensuremath{\mathrm{\,ke\kern -0.1em V}}\xspace}
\newcommand{\ev}{\ensuremath{\mathrm{\,e\kern -0.1em V}}\xspace}
\newcommand{\gevc}{\ensuremath{{\mathrm{\,Ge\kern -0.1em V\!/}c}}\xspace}
\newcommand{\mevc}{\ensuremath{{\mathrm{\,Me\kern -0.1em V\!/}c}}\xspace}
\newcommand{\gevcc}{\ensuremath{{\mathrm{\,Ge\kern -0.1em V\!/}c^2}}\xspace}
\newcommand{\gevgevcccc}{\ensuremath{{\mathrm{\,Ge\kern -0.1em V^2\!/}c^4}}\xspace}
\newcommand{\mevcc}{\ensuremath{{\mathrm{\,Me\kern -0.1em V\!/}c^2}}\xspace}

\usepackage{cite} 
\usepackage{mciteplus}

\begin{document}
\maketitle
\begin{abstract}
The LHCb Collaboration recently reported a measurement of the difference in direct CP asymmetries for the decays $\Lambda_b^0 \to J/\psi \, p \, h^-$ (with $h = K, \pi$), offering new experimental constraints on the decay dynamics of heavy baryons into charmonium final states. Inspired by these findings, we explore the branching ratios and direct CP violations for the decays $\Lambda_b^0 \to P_c^+(4312, 4440, 4457)\,h^-$ within the framework of final-state rescattering. Our analysis indicates that the branching fractions for $\Lambda_b^0 \to P_c^+ \pi^-$ lie around the $10^{-6}$ level, with the corresponding direct CP asymmetries approaching approximately $1\%$. In contrast, the direct CP violation for the decay $\Lambda_b^0 \to P_c^+ K^-$ is found to be very small, while its branching ratios show a strong dependence on the spin assignments of the $P_c$ states. These predictions may provide useful guidance for more precise CP measurements and amplitude analyses in the $P_c$ region in future experiments.
\newpage
\end{abstract}
\section{Introduction}

CP violation is one of the essential building blocks for understanding the dynamical origin of the baryon-antibaryon asymmetry observed in the Universe. In 2025, the LHCb Collaboration reported the first confirmation of CP asymmetry in the baryon decay via the multibody decaying channel $\Lambda^{0}_{b}\to pK^{-}\pi^{+}\pi^{-}$ with a significance of $5.2\sigma$\cite{LHCb:2025ray}. Specifically, the experimental studies reported  localized CP asymmetries across different phase space regions of the Dalitz plot. These experimental measurements are either consistent with theoretical prediction based on the $N\pi\to N\pi\pi$ rescatterings\cite{Wang:2024oyi}, or interpreted via final state interaction calculations with the hadron scattering framework\cite{Shang:2026knt,Feng:2026soj}. It opens a new avenue for studying QCD dynamics through flavor CP violation, offering opportunities for precision tests of the Standard Model and future constraints on possible new physics. 

Recently, the LHCb Collaboration measured the CP asymmetry difference $\Delta A_{CP}$ in the charmonium-mediated decays $\Lambda^{0}_{b}\to J/\psi  ph^{-}$ ($h=K,\pi$). The observable defined as $\Delta A_{CP} \equiv A_{CP}(\Lambda_b^0 \to J/\psi\, p\, \pi^-)\;-\;A_{CP}(\Lambda_b^0 \to J/\psi\, p\, K^-)$, was determined to be $(4.31 \pm 1.06 \pm 0.28)\%$\cite{LHCb:2025mnp}. The decay mode $\Lambda^{0}_{b}\to J/\psi\,pK^{-}$ proceeds through the quark level transition $b\to c\bar{c}s$, in which the relevant weak phases are nearly aligned. As a result, the Standard Model predicts a highly suppressed direct CP asymmetry in this channel. In contrast, the decay $\Lambda^{0}_{b}\to J/\psi\,p\pi^{-}$ proceeds via $b\to c\bar{c}d$, which involves a sizable weak phase difference and can therefore exhibit a potentially significant CP asymmetry. Under the assumption that $A_{CP}(\Lambda^{0}_{b}\to J/\psi\,pK^{-})$ is negligibly small, a nonzero $\Delta A_{CP}$ becomes a sensitive probe of CP violation in $\Lambda^{0}_{b}\to J/\psi\,p\pi^{-}$. Notably, the Dalitz analysis presented in \cite{LHCb:2025mnp} shows a comparatively larger asymmetry in the kinematic region where $J/\psi\,p$ invariant mass lies near the pentaquark states $P_{c}$. This observation suggests that the region may provide a particularly promising window for probing direct CP asymmetry in the decay $\Lambda^{0}_{b}\to P_{c}^+\pi^{-}$.  Such a measurement, if experimentally confirmed, would constitute the first observation of CP violation in a decay involving the pentaquark states. Motivated by this prospect, we investigate CP asymmetry in $\Lambda^{0}_{b}\to P_{c}^+h^{-}$ in the present work.

 In recent years, CP violation in baryon decays has been studied extensively\cite{Lu:2009cm,Hsiao:2014mua,Hsiao:2017tif,Zhu:2016bra,Zhang:2015wba,Zhang:2021fdd,Zhang:2025mne,Wang:2024qff,Shen:2023eln,Han:2024kgz,Wang:2024oyi,Han:2025tvc,He:2025msg,Chen:2025puj,Zhang:2025jnw,Yu:2025ekh,Duan:2024zjv,Feng:2026soj,Shang:2026knt,Jia:2024pyb}. For example, recent pQCD calculations indicate that the mechanism of CP violation in baryon decays differs markedly from that in meson decays: a partial cancellation between the \(S\)- and \(P\)-wave contributions suppresses the direct CP asymmetry. This mechanism provides a new dynamical interpretation of the currently small CPV observed in two-body \(\Lambda_b\) decays. However, for processes involving pentaquark final states, short-distance factorization approaches such as pQCD/QCDF generally require nonperturbative inputs, in particular light-cone distribution amplitudes (LCDAs). At present, these inputs for pentaquark states are still poorly constrained, so fully reliable quantitative predictions within factorization frameworks remain challenging. In this context, the final-state-interaction (FSI) approach is more suitable for the present work, since it can effectively incorporate long-distance hadronic dynamics and strong phases without relying on pentaquark LCDAs.

FSI has been successfully applied to nonleptonic decays of heavy hadrons. In this framework, long-distance amplitudes are treated at the hadron level, allowing a systematic estimate of nonfactorizable contributions to topological amplitudes \cite{Cheng:2004ru,Wang:2020gmn}. For CP-violation mechanisms, FSI provides a picture different from QCD factorization: the strong phase is generated primarily by hadronic rescattering in the final state, rather than by quark--antiquark loop effects (the BSS mechanism) \cite{Bander:1979px}. By introducing quasi-two-body intermediate processes involving resonant states, FSI also offers a practical computational strategy for multibody decays. Studies of two-body B-meson final states already showed that rescattering can be crucial for both branching fractions and direct CP asymmetries, and that the absorptive part of the full amplitude can be extracted by combining short-distance amplitudes with long-distance scattering amplitudes via the optical theorem \cite{Cheng:2004ru}. In recent years, this approach has been extended from mesonic to baryonic systems\cite{Li:2020qrh, Jiang:2018oak, Li:2018epz, Han:2021gkl, Han:2021azw, Yu:2017zst, Jia:2024pyb,Feng:2026soj,Shang:2026knt,Duan:2024zjv,Hu:2025pjg,Hu:2024uia}. In Ref. \cite{Jia:2024pyb}, charmed-baryon decays were studied in the rescattering framework using a loop-integration formalism in place of the optical theorem; with only one model parameter, the branching fractions of eight channels can be described simultaneously. Applied further to bottom-baryon decays, the same framework with two model parameters accounts for five branching fractions and CP-violating observables \cite{Duan:2024zjv}. In addition, FSI can naturally accommodate processes involving excited states\cite{Feng:2026soj,Shang:2026knt}. In this context, at the phenomenological level, it provides important guidance for experimental measurements and preliminary theoretical investigations, despite suffering from systematic uncertainties that are difficult to improve further. Here, we extend the approach to the decays $\Lambda^{0}_{b}\to P_{c}^+h^{-}$.

On the other hand, the LHCb Collaboration  has discovered three  hidden-charm pentaquark candidates
$P_c(4312)^+$, $P_c(4440)^+$, and $P_c(4457)^+$ in the $J/\psi\,p$ invariant mass spectrum of the 
$\Lambda_b^0\to J/\psi\,pK^-$ decay~\cite{LHCb:2019kea}.    
Various theoretical scenarios were proposed to reveal  the nature of these pentaquark states,
including hadronic molecules\cite{Chen:2019asm,He:2019ify,Chen:2019bip,Xiao:2019aya,Sakai:2019qph,Yamaguchi:2019seo,He:2019rva,Liu:2019zvb,PavonValderrama:2019nbk,Meng:2019ilv,Du:2019pij,Ling:2021lmq,Dong:2021juy,Ozdem:2021ugy,Pan:2022xxz,Zhang:2023czx,Pan:2022whr,Liu:2023wfo,Pan:2023hrk}, compact pentaquark \cite{Ali:2019npk,Wang:2019got,Cheng:2019obk,Weng:2019ynv,Zhu:2019iwm,Pimikov:2019dyr,Ruangyoo:2021aoi}, baryocharmonium\cite{Eides:2019tgv},
and kinematical effects\cite{Nakamura:2021qvy,Burns:2022uiv}. Among the various theoretical interpretations, the hadronic molecular picture is the most favored. This is because both the masses and the mass splittings of these pentaquark states can be naturally accommodated within the heavy‑quark spin symmetry (HQSS) multiplet framework~\cite{Liu:2019zvb}. In this picture,  the $P_c(4312)^+$ is  interpreted  as  a
$\Sigma_c\bar D$ bound state with $J^P=\frac12^-$, while the   $P_c(4440)^+$ and $P_c(4457)^+$ are treated  as
  $\Sigma_c\bar D^{*}$ bound states with  either  $J^P=\frac12^-$ or $\frac32^-$. 
However, the spin-parity quantum numbers of the $P_c(4440)^+$ and $P_c(4457)^+$ have not yet been determined experimentally.  While a number of theoretical studies have proposed various physical observables to discriminate between possible spin assignments for these states, no promising observable has yet been established.
In  the two body decays $\Lambda_b^0\to P_c^+ h^-$,  CP violations may offer new insights into the the internal structures of three pentaquark states and spin order of $P_c(4440)^+$ and $P_c(4457)^+$.  Therefore, we will compute the branching fractions and direct CP violation of the two-body decays $\Lambda_b^0\to P_c^+ h^-$  assuming the three pentaquark states as the hadronic molecules.

The paper is organized as follows. In Sec. II, we concisely introduce the theoretical framework. Since many details of the computational methodology have already been presented in our previous work~\cite{Duan:2024zjv}, we focus here only on new aspects. In Sec. III,  we present the numerical results for the helicity amplitudes, branching ratios, and CP asymmetries. Finally, we provide a summary in Sec. IV. The effective Lagrangian and the analytical expressions for the long-distance amplitudes are collected in the Appendixes.

\section{Theoretical framework}\label{framework}

\begin{figure}[H]
\centering 
     \includegraphics[width=15cm,height = 7.5cm]
     {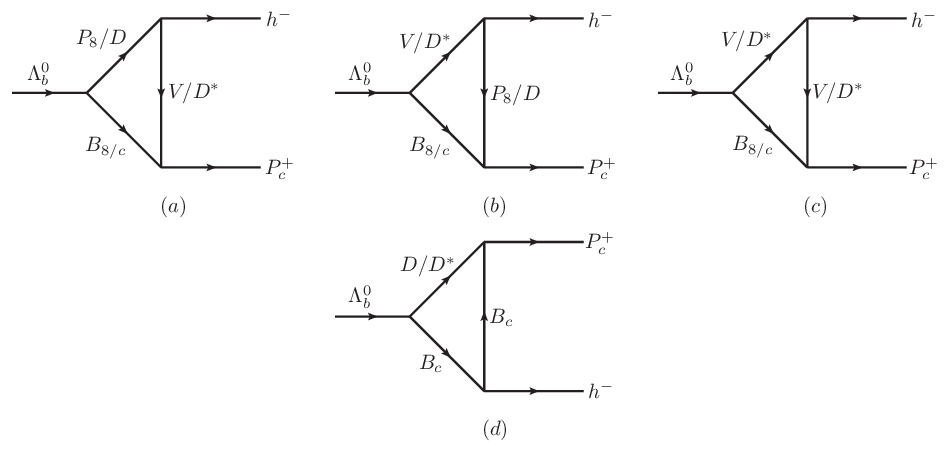}
\caption{The triangle diagrams of $\Lambda^{0}_{b}\to P_{c}^+h^{-}$ considered in this work, where the $B_{c},B_{8}$ represent charmed and octet light baryons, and $P,V,D,D^{*}$ are pseudoscalar and vector octet, $D$ and $D^{*}$ mesons, respectively. (a) represents rescattering between a pseudoscalar meson and a baryon via vector-meson exchange, (b) represents rescattering between a vector meson and a baryon via pseudoscalar-meson exchange, (c) represents rescattering between a vector meson and a baryon via vector-meson exchange, and (d) represents rescattering between a charmed meson and a charmed baryon via charmed-baryon exchange.}\label{triangle diagrams of Lb to Pc Pf}
\end{figure}

 Under final states interaction approach, the decays $\Lambda^{0}_{b}\to P_{c}^+h^{-}$  can be viewed  as a two-step process: first, the $\Lambda^{0}_{b}$ decays into a pair of intermediate hadrons, which then undergo rescattering via the exchange of a single hadron, ultimately transforming into the final states of concern.  Within this framework, the underlying physical picture is described by the triangle diagrams illustrated in Fig. \ref{triangle diagrams of Lb to Pc Pf}. In this work, the triangle loops incorporate the pseudoscalar-meson octet $P_8$, the baryon octet $B_8$, the charmed mesons $D^{(*)}$, as well as the charmed baryons $B_c$(i.e., the antitriplet $B_{\bar{3}}$ and the sextet $B_6$).

To calculate the triangle diagrams above, we treat the weak vertices within the naive factorization approach, while the Feynman rules for the strong vertices are derived from effective hadronic Lagrangians. The former have been introduced comprehensively in our previous work\cite{Duan:2024zjv}. As an illustration, we consider diagram (a) in Fig.\ref{triangle diagrams of Lb to Pc Pf} as an example. The corresponding amplitude is expressed as an integral over the internal momentum $k$:  
\begin{equation}
 \begin{aligned}
\mathcal{M}[D,B_c, D^{*};P_c^{1/2^-}]
     &=\int \frac{d^4k}{(2\pi)^4} g_{D^{*}DP}\,g_{ D^* \Lambda_c^+ P_{c}}^{1/2^{-}}\,p_{3\alpha} \,\bar{u}(p_4)\gamma^\nu \gamma_5(g_{\mu \nu}-\frac{p_{4\mu}p_{4\nu}}{m_{P_c}^2})(-g^{\alpha \mu}+\frac{k^\alpha k^\mu}{m_k^2}) \\
    &\times(\not\! p_2 +m_2)(A+B\gamma_{5})u(p_i)
    \frac{\mathcal{F}}{(p_1^2-m_1^2)(p_2^2-m_2^2)(k^2-m_k^2)}\,.
         \end{aligned}
\end{equation}
The momenta $p_i$, $p_3$, and $p_4$  represent those of the initial baryon \(\Lambda_b^0\), the final-state meson \(h^-\), and the pentaquark \(P_c\), respectively, while \(p_1\), \(p_2\), and \(k\) denote the momenta of the intermediate particles. Here, the notation $\mathcal{M}[D, B_c, D^{*}; P_c^{1/2^-}]$ denotes the triangle amplitude, in which $D$ and $B_c$ are the intermediate particles produced at the weak vertex, $D^{*}$ is the exchanged particle in the scattering loop, and $P_c^{1/2^-}$ is the final pentaquark state with spin parity $J^{P}=1/2^{-}$.

The effective form factor $\mathcal{F}$ follows the same parametrization as in our previous work: $\mathcal{F}(\Lambda, m_k) = \Lambda^4/((k^2 - m_k^2)^2 + \Lambda^4),$ where $\Lambda$ is a model parameter\cite{Duan:2024zjv}. For the decay $\Lambda_b^0 \to P_c^+ h^-$, the rescattering processes $\Lambda_c^+ D^-_{(s)} \to P_c^+ h^-$ and $p\pi^-(K^-) \to P_c^+ h^-$ involve different intermediate particles and hence exhibit distinct off‑shell effects in the loop. Accordingly, we introduce two model parameters, $\Lambda_{\rm charm}$ and $\Lambda_{\rm charmless}$, to characterize these two distinct rescattering modes, following the method used in our previous work on two-body charmless nonleptonic $\Lambda_b^0$ decays\cite{Duan:2024zjv}. Recently, the LHCb collaboration performed a binned Dalitz-plot analysis of the direct $CP$ asymmetry in $\Lambda_b^0\to pK_S^0\pi^-$ and found that $A_{CP}$ in the phase space region dominated by $K^{*}(892)^-$ approaches zero, indicating small $CP$ asymmetry in this channel \cite{LHCb:2025ozp}—consistent with our theoretical prediction \cite{Duan:2024zjv}. It largely boosts our confidence, and hence we adopt similar values for the model parameters $\Lambda$ in the decays $\Lambda_b^0 \to P_c^+ h^-$. Finally, the full amplitude expressions for all triangle diagrams are collected in Appendix~\ref{app.B}.

\section{Numerical results and discussions}
\subsection{Input parameters}
The global input parameters used in this work are summarized as follows. The masses of the three final-state pentaquark states $P_c^+$ are taken as 4.312 GeV, 4.440 GeV, and 4.457 GeV, respectively. The baryon and meson masses are $m_{\Lambda_b^0} = 5.619~\mathrm{GeV},  m_{\Lambda_c^+} = 2.286~\mathrm{GeV},  m_p = 0.938~\mathrm{GeV}, \
m_{D} = 1.869~\mathrm{GeV},  m_{\pi} = 0.140~\mathrm{GeV}, m_{D^*} = 2.010~\mathrm{GeV}, m_\rho = 0.770~\mathrm{GeV}$. The quark masses are current masses with values $
m_u = 2.16~\mathrm{MeV}, m_d = 4.70~\mathrm{MeV}, m_s = 93.5~\mathrm{MeV},m_c = 1.27~\mathrm{GeV},m_b = 4.18~\mathrm{GeV},$ from Ref.~\cite{ParticleDataGroup:2024cfk}. The CKM matrix elements are adopted at leading order of Wolfenstein parameterization with $A=0.823, \rho=0.141,\eta=0.349$ and $\lambda=0.225$ \cite{ParticleDataGroup:2024cfk}. The heavy-to-light transition form factors for $\Lambda^{0}_{b} \to p$ and $\Lambda^{+}_{c}$ are taken from Ref. \cite{Zhu:2018jet}. The decay constants are $f_{\pi} = 130.3$ and $f_{D} = 212$ MeV for pseudoscalar mesons, and $f_{\rho} = 216$ and $f_{D^*} = 230$ MeV for vector mesons \cite{Zhu:2018jet}.  The couplings of the pentaquark states to charmed baryons and mesons, as well as to protons and light mesons, are extracted from their partial widths, as shown in Table IV of Ref.~\cite{Lin:2019qiv}. The pentaquark-state couplings used in this work are summarized in Table \ref{coupling constants}. The remaining couplings can be found in our previous work\cite{Duan:2024zjv}.

\begin{table}[H]
\centering
\caption{The values of the pentaquark states couplings to  their constituents. The coupling constants are dimensionless. }
\label{coupling constants}
\begin{tabular}{ c c c c}
 \hline         
  \hline  
            & $P_c^+(4312)$ &$P_c^+(4440)$&$P_c^+(4457)$  \\
 \hline          
 $g_{P_c^{1/2}\pi N}$ & 0.1185 & 0.0402 & 0.0238  \\
$g_{P_c^{3/2}\pi N}$&  & 0.0006& 0.0003 \\ 
$g_{P_c^{1/2}\Lambda_cD}$ &0.0775&0.2019&  0.1714\\
$g_{P_c^{3/2}\Lambda_cD}$ &  &0.9179&  0.8528\\
$g_{P_c^{1/2}\Lambda_cD^{*}}$ &0.4565&0.2938&  0.2655\\
$g_{P_c^{3/2}\Lambda_cD^{*}}$ &&0.3753&  0.3675\\
$g_{P_c^{1/2}\rho N}$ & 0.0009 & 0.0185 & 0.0092  \\
$g_{P_c^{3/2}\rho N}$ &  & 0.0278 & 0.0160 \\
$g_{P_c^{1/2}\Sigma_cD}$ &1.8913&0.2744&  0.2411\\
$g_{P_c^{3/2}\Sigma_cD}$ &&1.9379&  2.0067\\
$g_{P_c^{1/2}\Sigma_cD^{*}}$ &&1.3581&  0.9252\\
$g_{P_c^{3/2}\Sigma_cD^{*}}$ &&2.3522&  1.6024\\
$g_{P_c^{1/2} \omega p}$ &0.0017&0.0361&  0.0208\\
$g_{P_c^{3/2} \omega p}$ &&0.0560&  0.0322\\
 \hline  
  \hline  
\end{tabular}
\end{table}

\subsection{Numerical results}

Using the model parameters $\Lambda_{\text{charm}}=1.0 \,\text{GeV}$ and $\Lambda_{\text{charmless}}=0.5 \,\text{GeV}$, we present the helicity amplitudes for the decays $\Lambda_b^0 \to P_c^+ \pi^-$ and $\Lambda_b^0 \to P_c^+ K^-$ in Tables \ref{Helicity Lb to Pc pi} and \ref{Helicity Lb to Pc K}, respectively.   These amplitudes can be directly tested in future experiments through partial‑wave analyses. The notation follows that of our previous work. For instance, $\mathcal{NC}$ $(\mathcal{C})$ denotes the total amplitude arising from light‑hadron (charmed‑hadron) scattering. We emphasize that the helicity amplitudes listed in Tables \ref{Helicity Lb to Pc pi} and \ref{Helicity Lb to Pc K} do not include CKM factors. The CKM symbols shown in the tables serve only to distinguish the different contributions.

\begin{sidewaystable}
\caption{Helicity amplitudes of $\Lambda_b^0 \to P_c^+\pi^-(10^{-8})$ with different $CKM$ factors}
 \renewcommand{\arraystretch}{1.5} 
\centering
\begin{tabular}{ccccccc}
\toprule
\toprule
 Decay modes & $H_{-\frac{1}{2}} ~(V_{ub}V_{ud}^*) $ &$H_{-\frac{1}{2}} ~(V_{cb}V_{cd}^*)$ &$H_{-\frac{1}{2}} ~(V_{tb}V_{td}^*)$ & $H_{\frac{1}{2}} ~(V_{ub}V_{ud}^*) $ &$H_{\frac{1}{2}}~(V_{cb}V_{cd}^*)$ & $H_{\frac{1}{2}} ~(V_{tb}V_{td}^*)$ \\
\midrule
$\mathcal{NC}(\Lambda_b^0 \to P_c^{1/2}(4312)\pi^-)$& -4.251 + 2.360 $i$ &... & -0.081 + 0.052 $i$& -2.367 - 2.471 $i$&...&-0.044 - 0.054 $i$\\
$\mathcal{C}(\Lambda_b^0 \to P_c^{1/2}(4312)\pi^-)$& ... &-69.983 - 318.344 $i$ & 10.812 - 12.316 $i$ & ...&-382.465 + 246.552 $i$& - 5.204 + 13.197 $i$\\

\midrule 
$\mathcal{NC}(\Lambda_b^0 \to P_c^{1/2}(4440)\pi^-)$& 1.983 - 0.691 $i$ &... & 0.334 - 0.028 $i$& - 10.926 - 0.447 $i$&...&-0.060 + 0.006 $i$\\
$\mathcal{C}(\Lambda_b^0 \to P_c^{1/2}(4440)\pi^-)$& ... &140.611 - 44.374 $i$ &7.102 - 1.664 $i$ & ...&-240.355 + 126.605 $i$& - 8.163 + 7.778 $i$\\

\midrule
$\mathcal{NC}(\Lambda_b^0 \to P_c^{3/2}(4440)\pi^-)$& -4.478 + 5.027  $i$ &... & -0.147 + 0.117  $i$& 2.139 - 2.156 $i$&...& -0.092 - 0.016 $i$\\
$\mathcal{C}(\Lambda_b^0 \to P_c^{3/2}(4440)\pi^-)$& ... &-226.499 + 200.560 $i$ &  -7.279 + 6.803  $i$ & ...&4.713 + 161.707  $i$& - 3.638 + 8.335  $i$\\

\midrule
$\mathcal{NC}(\Lambda_b^0 \to P_c^{1/2}(4457)\pi^-)$& 1.457 - 0.368 $i$ &... & 0.178 - 0.015 $i$& - 5.772 - 0.311 $i$&...&-0.036 + 0.001 $i$\\
$\mathcal{C}(\Lambda_b^0 \to P_c^{1/2}(4457)\pi^-)$& ... &134.438 - 40.664 $i$ & 6.649 - 1.436 $i$ & ...&-218.503 + 112.139   $i$& - 7.615 + 6.932  $i$\\

\midrule 
$\mathcal{NC}(\Lambda_b^0 \to P_c^{3/2}(4457)\pi^-)$&- 2.722 + 3.063 $i$ &... & -0.088 + 0.071 $i$& 1.129 - 1.304  $i$&...&-0.055 - 0.010 $i$\\
$\mathcal{C}(\Lambda_b^0 \to P_c^{3/2}(4457)\pi^-)$& ... &-217.008 + 182.601 $i$ & - 7.161 + 6.320 $i$ & ...&-5.088 + 141.572  $i$& -4.009 + 7.729  $i$\\
\bottomrule
\bottomrule
\end{tabular}\label{Helicity Lb to Pc pi}
\end{sidewaystable}

\begin{sidewaystable}
\caption{Helicity amplitudes of $\Lambda_b^0 \to P_c^+K^-(10^{-8})$ with different $CKM$ factors}
 \renewcommand{\arraystretch}{1.5} 
\centering
\begin{tabular}{ccccccc}
\toprule
\toprule
 Decay modes & $H_{-\frac{1}{2}} ~(V_{ub}V_{us}^*) $ &$H_{-\frac{1}{2}} ~(V_{cb}V_{cs}^*)$ & $H_{-\frac{1}{2}} ~(V_{tb}V_{ts}^*)$ & $H_{\frac{1}{2}} ~(V_{ub}V_{us}^*) $ &$H_{\frac{1}{2}} ~(V_{cb}V_{cs}^*)$ & $H_{\frac{1}{2}} ~(V_{tb}V_{ts}^*)$ \\
\midrule 
$\mathcal{NC}(\Lambda_b^0 \to P_c^{1/2}(4312)K^-)$& 0.131 - 0.088 $i$ &... & -0.013 - 0.001 $i$& 0.347 - 0.174 $i$&...&-0.003 - 0.004 $i$\\
$\mathcal{C}(\Lambda_b^0 \to P_c^{1/2}(4312)K^-)$& ... &66.239 - 368.781 $i$ & 2.122 - 14.156 $i$ & ...&-86.776 - 282.747 $i$&-3.794 + 15.176 $i$\\

\midrule 
$\mathcal{NC}(\Lambda_b^0 \to P_c^{1/2}(4440)K^-)$& 9.647 - 0.045 $i$ &... & -0.156 + 0.015 $i$& 9.887 - 0.543 $i$&...&-0.033 - 0.022 $i$\\
$\mathcal{C}(\Lambda_b^0 \to P_c^{1/2}(4440)K^-)$& ... &169.175 - 230.984 $i$ &6.391 - 8.942 $i$ & ...&-228.386 + 73.249 $i$&-9.268 + 6.007 $i$\\

\midrule 
$\mathcal{NC}(\Lambda_b^0 \to P_c^{3/2}(4440)K^-)$& 4.766 - 0.864 $i$ &... & 0.167 - 0.026 $i$& 3.216 - 0.454 $i$&...&0.228 - 0.040 $i$\\
$\mathcal{C}(\Lambda_b^0 \to P_c^{3/2}(4440)K^-)$& ... &-84.380 + 98.379 $i$ & - 4.629 + 5.176 $i$ & ...&-27.335 + 26.260  $i$& - 4.998 + 6.242  $i$\\

\midrule
$\mathcal{NC}(\Lambda_b^0 \to P_c^{1/2}(4457)K^-)$& 5.969 - 0.031 $i$ &... & -0.096 + 0.009 $i$& 6.112 - 0.332 $i$&...&-0.021 - 0.014 $i$\\
$\mathcal{C}(\Lambda_b^0 \to P_c^{1/2}(4457)K^-)$& ... &162.182 - 203.396 $i$ & 6.181 - 7.971 $i$ & ...&-211.658 + 67.265  $i$& - 8.668 + 5.326  $i$\\

\midrule
$\mathcal{NC}(\Lambda_b^0 \to P_c^{3/2}(4457)K^-)$& 2.717 - 0.496 $i$ &... & 0.096 - 0.015 $i$& 1.847 - 0.266 $i$&...&0.130 - 0.023 $i$\\
$\mathcal{C}(\Lambda_b^0 \to P_c^{3/2}(4457)K^-)$& ... &-85.416 + 95.409 $i$ & - 4.760 + 5.005 $i$ & ...&-28.593 + 25.610  $i$& -5.197 + 6.024  $i$\\
\bottomrule
\bottomrule
\end{tabular}\label{Helicity Lb to Pc K}
\end{sidewaystable}

The branching ratios and direct CP asymmetries are calculated by 
\begin{equation}
 \begin{aligned}
BR\left[\Lambda^{0}_{b}\to P_c^+\pi^{-}(P_c^+K^{-})\right]=\frac{|p_{c}|}{8\pi M^{2}_{\Lambda_{b}}\Gamma_{\Lambda^{0}_{b}}}\frac{1}{2}\left(\left|H_{+1/2}\right|^{2}+\left|H_{-1/2}\right|^{2}\right),~~~~~a^{dir}_{CP}=\frac{\Gamma-\bar{\Gamma}}{\Gamma+\bar{\Gamma}} \,,
         \end{aligned}
\end{equation}
where $p_{c}$ is the final $P_c^+$ momentum in the $\Lambda^{0}_{b}$ rest frame, and $1/2$ in $BR\left[\Lambda^{0}_{b}\to P_c^+\pi^{-}(P_c^+K^{-})\right]$ accounts for the initial spin average. Using  the model parameters  $\Lambda_{\text{charm}}=1.0\pm 0.1$~GeV and $\Lambda_{\text{charmless}}=0.5\pm 0.1$~GeV,  we present the  branching ratios and CP violations in Table \ref{BR and CPV of Pc pi} and Table~\ref{BR and CPV of Pc K}, respectively.

\begin{table}[H]
\centering
\renewcommand{\arraystretch}{1.4} 
\caption{The numerical results for the branching ratio and CP violation in $ \Lambda_b^0 \to P_c^+\pi^-$.}
\label{BR and CPV of Pc pi}
\begin{tabular}{ c c c }
  \hline  
 \hline          
 Decay modes  & BR$(10^{-6})$ & Direct CP($10^{-2}$) \\
 \hline          
 $\Lambda_b^0\to P_c^{1/2^-}(4312)\pi^-$  & $4.27^{+1.13}_{-0.98} $ & $0.13^{+0.08}_{-0.09}$\\
$\Lambda_b^0\to P_c^{1/2^-}(4440)\pi^-$  & $1.12^{+0.19}_{-0.14}$ & $-1.69^{+0.44}_{-0.28}$ \\
 $\Lambda_b^0\to P_c^{3/2^-}(4440)\pi^-$  &  $1.44^{+0.73}_{-0.49}$ & $0.72^{+0.43}_{-0.34}$\\
 $\Lambda_b^0\to P_c^{1/2^-}(4457)\pi^-$  & $0.94^{+0.13}_{-0.10}$ & $-1.18^{+0.26}_{-0.18}$\\
 $\Lambda_b^0\to P_c^{3/2^-}(4457)\pi^-$  & $1.21^{+0.57}_{-0.38}$ & $0.58^{+0.37}_{-0.30}$ \\
 \hline         
  \hline          
\end{tabular}
\end{table}

\begin{table}[H]
\centering
\caption{The numerical results for the branching ratio and CP violation in $ \Lambda_b^0 \to P_c^+K^-$. }
\label{BR and CPV of Pc K}
\begin{tabular}{ c c c }
 \hline
 \hline 
 Decay modes & BR$(10^{-6})$ &  Direct CP($10^{-2}$)   \\
 \hline          
 $\Lambda_b^0\to P_c^{1/2^-}(4312)K^-$\\
 This work  & $51.70^{+12.41}_{-10.80}$ & {$-6.71 \times 10^{-4}$}\\
 Previous work\cite{Pan:2023hrk}  &   ${35.18/98.88}$ &...\\
  \hline 
$\Lambda_b^0\to P_c^{1/2^-}(4440)K^-$ \\
This work  & $28.56^{+3.31}_{-2.81}$  & $0.05^{+0.01}_{-0.01}$\\
Previous work\cite{Pan:2023hrk} &   $15.30$  &...\\
 \hline 
 $\Lambda_b^0\to P_c^{3/2^-}(4440)K^-$ \\
 This work  &  $3.55^{+0.35}_{-0.33}$  & $-0.10^{+0.02}_{-0.01}$\\
 Previous work\cite{Pan:2023hrk}  &      $5.21$  &...         \\
  \hline 
 $\Lambda_b^0\to P_c^{1/2^-}(4457)K^-$\\
 This work  & $23.56^{+2.46}_{-2.08}$  & $0.03^{+0.005}_{-0.005}$ \\
  Previous work\cite{Pan:2023hrk}  &      $27.23$      &...   \\
  \hline 
 $\Lambda_b^0\to P_c^{3/2^-}(4457)K^-$\\
  This work  & $3.41^{+0.30}_{-0.29}$ &$-0.06^{+0.01}_{-0.01}$\\
  Previous work\cite{Pan:2023hrk}  &      $0.48$  &...        \\
 \hline    
 \hline  
\end{tabular}
\end{table}

\begin{table}[H]
\centering
\renewcommand{\arraystretch}{1.4} 
\caption{Ratio of branching fractions:$BR(\Lambda_b^0 \to P_c^+ \pi^-)/BR(\Lambda_b^0 \to P_c^+ K^-)$}
\label{BR Ratio}
\begin{tabular}{ c c c } 
 \hline
 \hline 
 BR ratio  &  Value      &  \\
 \hline         
 $\mathrm{BR}(\Lambda_b^0\to P_c^{1/2}(4312)\pi^-)/\mathrm{BR}(\Lambda_b^0\to P_c^{1/2}(4312)K^-)$  & $0.083^{+0.002}_{-0.002}$ & \\
$\mathrm{BR}(\Lambda_b^0\to P_c^{1/2}(4440)\pi^-)/\mathrm{BR}(\Lambda_b^0\to P_c^{1/2}(4440)K^-)$  & $0.039^{+0.002}_{-0.001}$ &  \\
$\mathrm{BR}(\Lambda_b^0\to P_c^{3/2}(4440)\pi^-)/\mathrm{BR}(\Lambda_b^0\to P_c^{3/2}(4440)K^-)$  &  $0.406^{+0.151}_{-0.109}$ & \\
 $\mathrm{BR}(\Lambda_b^0\to P_c^{1/2}(4457)\pi^-)/\mathrm{BR}(\Lambda_b^0\to P_c^{1/2}(4457)K^-)$ & $0.040^{+0.001}_{-0.001}$ & \\
$\mathrm{BR}(\Lambda_b^0\to P_c^{3/2}(4457)\pi^-)/\mathrm{BR}(\Lambda_b^0\to P_c^{3/2}(4457)K^-)$ & $0.355^{+0.125}_{-0.089}$ &  \\
 \hline
 \hline 
\end{tabular}
\end{table}

\subsection{Discussions}
Some discussions are in order:
\begin{itemize}

\item In Table \ref{BR and CPV of Pc pi}, we present the results of the CP asymmetries and branching ratios for the decay $\Lambda_b^0 \to P_c^+\pi^-$.  Notably, for the pentaquark states $P_c(4440)$ and $P_c(4457)$ with spin-parity $J^P=1/2^-$, the  CP asymmetries are estimated to be  of order $\mathcal{O}(1\%)$. These values not only differ in sign but also are larger in magnitude than those for the $J^P=3/2^-$ assignment. Therefore, we conclude that our predictions may help to determine the spin order of $P_c(4440)$ and $P_c(4457)$ if a nonzero CP asymmetry in $\Lambda_b^0 \to P_c^+\pi^-$ is determined by future experiments.
\item In Table \ref{BR and CPV of Pc K}, we present the results of the CP asymmetries and branching ratios for  the decay $\Lambda_b^0 \to P_c^+ K^-$. Our results are generally comparable with those obtained in our previous work~\cite{Pan:2023hrk} for most channels. However, the last entry in Table~\ref{BR and CPV of Pc K}, corresponding to $\Lambda_b^0\to P_c^{3/2}(4457)K^-$, shows a noticeable deviation from the previous result. This deviation mainly originates from the different values of the $P_c^{3/2}\Lambda_c D^*$ coupling adopted in the two analyses. Since this channel is dominated by the rescattering process $\Lambda_b^0\to \bar{D}_s^{(*)}\Lambda_c^+\to P_c K^-$, the change in this coupling input gives rise to a sizable difference in the predicted branching ratio.

Within the set of intermediate states considered in this work, our results indicate that pentaquark production in $\Lambda_b^0$ decays is mainly driven by the weak transitions $\Lambda_b^0\to \bar{D}_s^{(*)}\Lambda_c^+$. In principle, excited charmed-strange mesons and $\Lambda_c^+$ baryons may also contribute to the rescattering amplitudes. However, a systematic inclusion of such excited intermediate states would require additional inputs, such as the relevant weak transition form factors and the strong couplings entering the rescattering vertices, which are not well constrained at present. Consequently, a systematic treatment of these channels would introduce many additional input parameters and enlarge the theoretical uncertainties. To strike a balance between a more complete treatment and the controllability of theoretical uncertainties, the present study is restricted to these ground-state intermediate states. A more complete coupled-channel analysis including such excited states is left for future work.


For the pentaquark states $P_c(4440)$ and $P_c(4457)$, we find that the branching ratios for the $J^P=1/2^-$ assignment are approximately one order of magnitude  larger than those for the $J^P=3/2^-$ assignment. This difference could also help us to determine the spin order of $P_c(4440)$ and $P_c(4457)$ in the future experiment. The CP asymmetry for the decay  $\Lambda_b^0 \to P_c^+ K^-$ is very small since the helicity amplitudes in Table~\ref{Helicity Lb to Pc K} are dominated by a single component with CKM factor $V_{cb}V_{cs}^\ast$.

\item We also estimate the ratio $\mathrm{BR}(\Lambda_b^0\to P_c^+\pi^-)/\mathrm{BR}(\Lambda_b^0\to P_c^+K^-)$ and present the results in Table \ref{BR Ratio}. The relative decay rates were estimated only by assuming the dominance of the external W-emission amplitude in the Ref.~\cite{Cheng:2015cca} 
\begin{equation}
\frac{\mathrm{BR}(\Lambda_b^0\to P_c^+\pi^-)}{\mathrm{BR}(\Lambda_b^0\to P_c^+K^-)}
\approx 0.07\sim0.08,
\label{ratio_K_pi}
\end{equation}
which is consistent with our results for the spin-$1/2$ assignment.
It is noting that Eq.~\eqref{ratio_K_pi} should be viewed as a naive CKM power counting, which implicitly assumes the reduced hadronic amplitudes of the $\pi$ and $K$ channels are approximately the same.
This assumption is roughly valid for the $J^P=1/2^-$ assignment, but it breaks down for the $J^P=3/2^-$. In our rescattering model, the smallness of the $\Lambda_b^0\to P_c^{3/2}K^-$ branching ratios mainly originates from the destructive interference among the charmed triangle amplitudes. Such a strong cancellation however does not appear in the $P_c^{3/2}\pi$ channel. This difference is related to the different rescattering topologies in the two channels. As listed in Appendix~\ref{app.C}, within the dominant charmed rescattering contributions considered in this work, the $P_c^{3/2}\pi$ mode contains long-distance $\Sigma_c^{++}$- and $D^{(*)}$-exchange contributions, while the corresponding $P_c^{3/2}K$ mode is generated only through meson-exchange diagrams. Therefore, the amplitude of the $P_c^{3/2}K$ mode is relatively small, and the naive CKM realtion in Eq.~\eqref{ratio_K_pi} is no longer applicable.
In addition, for $\Lambda_b^0\to P_c^{3/2}h$, the lowest allowed partial wave is a P wave.  This effect further suppresses the $P_c^{3/2}K$ branching ratio since the $K$ channel has a smaller final-state momentum than the $\pi$ channel.

\item  Recently, the LHCb  Collaboration measured the CP-asymmetry difference $\Delta A_{CP}$ in $\Lambda_b^0\to J/\psi p h^-(h=\pi,K)$ decays\cite{LHCb:2025mnp}, providing evidence for a nonzero difference in direct CP asymmetries between the two decay modes. Owing to the different CKM structures, $\Lambda_b^0\to J/\psi p\pi^-$ is expected to be more likely to exhibit a larger direct CP asymmetry than $\Lambda_b^0\to J/\psi pK^-$. Furthermore, the paper presents the raw asymmetry $A_{\rm raw}$ across the two-dimensional Dalitz phase space of $\Lambda_b^0\to J/\psi p\pi^-$. In the region where $m^2(J/\psi p)$ is close to the known $P_c$ states, the contrast is enhanced, providing a potential window to search for CP violation associated with hidden-charm pentaquark production and to test our predictions for $\Lambda_b^0\to P_c h^-$. Our numerical results appear unable to fully account for the experimentally observed $\Delta A_{CP}$, suggesting the presence of additional contributions beyond those considered in our current framework. 
\end{itemize}

\section{Summary}

Within the final-state rescattering mechanism, we investigate the branching ratios and CP asymmetries for the decays $\Lambda_b^0 \to P_c^+\,h^-$ (with $h = K, \pi$), where the $P_c$ states refer to $P_{c}(4312)$, $P_{c}(4440)$, and $P_{c}(4457)$. Treating these pentaquark states as molecular candidates, we examine two possible spin-parity assignments for $P_{c}(4440)$ and $P_{c}(4457)$ as the $\bar{D}^*\Sigma_c$ molecules: either $J^P=1/2^-$ and $3/2^-$, or $J^P=3/2^-$ and $1/2^-$.  Our results indicate that the branching ratios of $\Lambda_b^0 \to P_c^+\pi^-$ are to be of order $10^{-6}$ and show little sensitivity to the spin assignment of the pentaquark states, while the corresponding  CP asymmetries are expected to be  order of  $\mathcal{O}(1\%)$. In contrast, the branching ratios for the decays $\Lambda_b^0 \to P_c^+K^-$ are strongly  dependent on the spin assignments. They reach at the $10^{-5}$ level for the $J^P=1/2^-$ assignment, but fall to the $10^{-6}$ for $J^P=3/2^-$. The CP asymmetries in this channel are found to be extremely small. We look forward to future experimental tests of our predictions, particularly regarding the CP violating for the decays  $\Lambda_b^0 \to P_c^+\pi^-$. Since the spin quantum numbers of the $P_c$ states  remain to be established, the physical observables computed in this work could provide new insights toward establishing them. 

\section{ACKNOWLEDGMENTS}
We thank Qin Qin for his valuable suggestions on the manuscript. This work was supported by the National Natural Science Foundation of China under Grants No. 12575086, No. W2543006, No. 12075126, No. 12335003, and No. 12247101, the Fundamental Research Funds for the Central Universities under Grants No. lzujbky-2024-oy02 and No. lzujbky-2025-jdzx07, the Natural Science Foundation of Gansu Province under Grant No. 25JRRA799, and the  “111 Center” under Grant No. B20063.  This work was also
 supported by the China Postdoctoral Science Foundation under Grant No. GZC20261850.
\appendix
\section{Effective Lagrangian}\label{app.A}

\begin{itemize}
    \item The effective Lagrangians for the couplings of the pentaquark states to baryon–meson channels are taken from Refs. \cite{Lu:2015fva, Lin:2019qiv,Pan:2023hrk}.

\begin{equation}
    \begin{split}
      \mathcal{L}_{P_{c}N\pi }^{1/2^{-}}&=g_{P_{c}N\pi}^{1/2}\bar{N}\vec{\tau}\cdot\vec{\pi} P_{c}+\mathrm{H.c.}\,,\\ 
    \mathcal{L}_{P_{ c}N\pi}^{3/2^{-}}&=\frac{g^{3/2}_{P_{ c} N \pi}}{m_{\pi}^{2}} \bar{N}\gamma_{ 5}\gamma_{ \mu}\vec{ \tau} \cdot \partial^{\mu} \partial_{ \boldsymbol{\nu}}\vec{ \pi} P_{ c}^{ \boldsymbol{\nu}} + \mathrm{H.c.}\,,\\
     \mathcal{L}^{1/2^-}_{P_{c}N  \rho}& = g_{P_{c}N  \rho}^{1/2}\bar N  \gamma _{5} \left(g_{\mu\nu}- \frac{p_{\mu}p_{\nu}}{m_{P_{c}}^{2}} \right) \gamma^{\nu}P_{c}\rho^{\mu} + \mathrm{H.c.}\,,
    \\ \mathcal{L}^{3/2^-}_{P_{c}N  \rho}& = g_{P_{c}N  \rho}^{3/2}\bar N P_{c\,\mu}\rho^{\mu} + \mathrm{H.c.}\,,\\
    \mathcal{L}^{1/2^-}_{P_{c}\Lambda_{c}\bar{D}}& = g_{P_{c}\Lambda_{c}\bar{D}}^{1/2}P_{c}\Lambda_{c} \bar{D} + \mathrm{H.c.}\,,\\
      \mathcal{L}_{P_c \Lambda_c\bar{D}}^{3/2^-}&=\frac{g^{3/2}_{P_c \Lambda_c\bar{D}}}{m_{\bar{D}}^{2}} \,\bar \Lambda_c \gamma_5 \gamma_\mu \partial^\mu \partial^\nu \bar{D} P_{c\,\nu}^+ + \mathrm{H.c.}\,,\\
      \mathcal{L}^{1/2^-}_{P_{c}\Lambda_{c}\bar{D}^{*}}& = g_{P_{c}\Lambda_{c}\bar{D}^{*}}^{1/2}\bar{\Lambda}_{c} \gamma _{5} \left(g_{\mu\nu}- \frac{p_{\mu}p_{\nu}}{m_{P_{c}}^{2}} \right) \gamma^{\nu}P_{c}D^{*\mu} + \mathrm{H.c.}\,,
    \\ \mathcal{L}^{3/2^-}_{P_{c}\Lambda_{c}\bar{D}^{*}}& = g_{P_{c}\Lambda_{c}\bar{D}^{*}}^{3/2}\bar{\Lambda}_{c}P_{c\,\mu}D^{*\mu} + \mathrm{H.c.}\,.
    \end{split}
\end{equation}

\item The Lagrangians involving $D^{(*)}$ mesons, pseudoscalar meson octet $P_8$ and vector meson octet $V$ :
\begin{equation}
 \begin{split}
  \mathcal{L}_{V P_8 P_8}
    &=\frac{\mathrm{i} g_{\rho\pi\pi}}{\sqrt{2}} \operatorname{Tr}\left[V^\mu\left[P_8, \partial_\mu P_8\right]\right]
    \,,\\
     \mathcal{L}_{V V P_8}
    &=\frac{4 g_{V V P_8}}{f_{P_8}} \varepsilon^{\mu \nu \alpha \beta} \operatorname{Tr}\left(\partial_\mu V_\nu \partial_\alpha V_\beta P_8\right)
    \,,\\
    \mathcal{L}_{D^\ast DP_8}
    &=-i g_{D^* D P_8}\left(D^i \partial^\mu P_{8 i j} D_\mu^{* j \dagger}-D_\mu^{* i} \partial^\mu P_{8 i j} D^{j \dagger}\right)\,,\\
    \mathcal{L}_{D^\ast D^\ast P_8}
    &=\frac{1}{2} g_{D^* D^* P_8} \varepsilon_{\mu \nu \alpha \beta} D_i^{* \mu} \partial^\nu P^{i j}_8 \overleftrightarrow{\partial}^\alpha D_j^{* \beta \dagger}\,,\\
     \mathcal{L}_{P_8 {B}_6 {B}_{\bar{3}}}
     &=g_{P_8 {B}_6 {B}_{\bar{3}}} \operatorname{Tr}\left[\bar{{B}}_6 i \gamma_5 P_8 {B}_{\bar{3}}\right]+H . c .\,.
 \end{split}
 \end{equation}

\end{itemize}

The matrices under SU(3) flavor group representations are given:

  \begin{equation}
 \begin{aligned}
P=\left(\begin{array}{ccc}
\frac{\pi^0}{\sqrt{2}}+\frac{\eta}{\sqrt{6}} & \pi^{+} & \mathrm{K}^{+} \\
\pi^{-} & -\frac{\pi^0}{\sqrt{2}}+\frac{\eta}{\sqrt{6}} & \mathrm{~K}^0 \\
\mathrm{~K}^{-} & \bar{K}^0 & -\sqrt{\frac{2}{3}} \eta
\end{array}\right), \quad {B}_6=\left(\begin{array}{ccc}
\Sigma_{\mathrm{c}}^{++} & \frac{1}{\sqrt{2}} \Sigma_{\mathrm{c}}^{+} & \frac{1}{\sqrt{2}} \Xi_{\mathrm{c}}^{\prime+} \\
\frac{1}{\sqrt{2}} \Sigma_{\mathrm{c}}^{+} & \Sigma_{\mathrm{c}}^0 & \frac{1}{\sqrt{2}} \Xi_{\mathrm{c}}^{\prime 0} \\
\frac{1}{\sqrt{2}} \Xi_{\mathrm{c}}^{\prime+} & \frac{1}{\sqrt{2}} \Xi_{\mathrm{c}}^{\prime 0} & \Omega_{\mathrm{c}}
\end{array}\right)\nonumber,
\end{aligned}
 \end{equation}

   \begin{equation}
 \begin{aligned}
V=\left(\begin{array}{ccc}
\frac{\rho^0}{\sqrt{2}}+\frac{\omega}{\sqrt{2}} & \rho^{+} & \mathrm{K}^{*+} \\
\rho^{-} & -\frac{\rho^0}{\sqrt{2}}+\frac{\omega}{\sqrt{2}} & \mathrm{~K}^{* 0} \\
\mathrm{~K}^{*-} & \bar{\mathrm{K}}^{* 0} & \phi
\end{array}\right), ~~~~~~~~~~~~~\quad {B}_{\bar{3}}=\left(\begin{array}{ccc}
0 & \Lambda_{\mathrm{c}}^{+} & \Xi_{\mathrm{c}}^{+} \\
-\Lambda_{\mathrm{c}}^{+} & 0 & \Xi_{\mathrm{c}}^0 \\
-\Xi_{\mathrm{c}}^{+} & -\Xi_{\mathrm{c}}^0 & 0
\end{array}\right)\nonumber,
\end{aligned}
 \end{equation}
 
  \begin{equation}
 \begin{aligned}
 {B_8}=\left(\begin{array}{ccc}
\frac{\Sigma^0}{\sqrt{2}}+\frac{\Lambda}{\sqrt{6}} & \Sigma^{+} & p \\
\Sigma^{-} & -\frac{\Sigma^0}{\sqrt{2}}+\frac{\Lambda}{\sqrt{6}} & n \\
\Xi^{-} & \Xi^0 & -\frac{2}{\sqrt{6}} \Lambda
\end{array}\right)
 ,~~~~~~~~~~~~~~~\quad D=\left(
\begin{matrix}
    D^0,D^+,D_s^+
\end{matrix}\right)\,\nonumber.
\end{aligned}
 \end{equation}


\section{Amplitudes of triangle diagram}\label{app.B}
The amplitudes of $\Lambda^{0}_{b} \rightarrow P_c^{1/2^-}h^- $:
\begin{equation}
	\begin{aligned}
    M[V,B_8,P_8]&=\int\frac{d^4k}{(2\pi)^4} g_{ P_8B_8P_{c}}^{1/2^{-}}\, g_{VP_8P_8}\bar u(p_4)(\not {p_2}+m_2)(-g^{\mu\nu}+\frac{p_1^\mu p_1^\nu}{m_1^2})(A_1\gamma_\mu\gamma_5+A_{2} \frac{p_{2\mu}}{m_{i}} \gamma_{5}+B_{1} \gamma_{\mu}+B_{2} \frac{p_{2\mu}} {m_{i}})\\&u(p_i)(p_{3\nu}-k_\nu)\frac{i^3 \mathcal{F}}{(p_1^2-m_1^2)(p_2^2-m_2^2)(k^2-m_k^2)}\,,
      	\end{aligned}
\end{equation}
\begin{equation}
	\begin{aligned}
    M[D^{*},B_c,\bar D]
    &=\int\frac{d^4k}{(2\pi)^4}g_{ D B_c P_{c}}^{1/2^{-}}\, g_{D^* D P_8}\,\bar u(p_4)(\not p_2+m_2)\,p_3^\alpha (-g_{\mu \alpha}+\frac{p_{1\mu}p_{1\alpha}}{m_1^2})(A_1\gamma_\mu\gamma_5+A_{2} \frac{p_{2\mu}}{m_{i}} \gamma_{5}+B_{1} \gamma_{\mu}+B_{2} \frac{p_{2\mu}} {m_{i}})\\&u(p_i)\frac{i^3 \mathcal{F}}{(p_1^2-m_1^2)(p_2^2-m_2^2)(k^2-m_k^2)}\,,
    \end{aligned}
\end{equation}
\begin{equation}
	\begin{aligned}
    M[D^{*},B_c,\bar D^{*}]
   &=\int\frac{d^4k}{(2\pi)^4}(-1)g_{ D^* B_c P_{c}}^{1/2^{-}} \,g_{D^{*}D^*P_8}\,\bar{u}(p_4)\gamma^\nu \gamma_5(g_{\mu \nu}-\frac{p_{4\mu}p_{4\nu}}{m_{P_c}^2})(\not p_2+m_2)(-g_{\mu \beta}+\frac{k_\mu k_\beta}{m_k^2})(-g_{nt}+\frac{p_{1n}p_{1t}}{m_1^2})\\&\varepsilon^{mn\alpha\beta}\,p_{1m}\,k_{\alpha}(A_1\gamma_t\gamma_5+A_{2} \frac{p_{2t}}{m_{i}} \gamma_{5}+B_{1} \gamma_{t}+B_{2} \frac{p_{2t}} {m_{i}})u(p_i)\frac{i^3 \mathcal{F}}{(p_1^2-m_1^2)(p_2^2-m_2^2)(k^2-m_k^2)}\,,
    \end{aligned}
\end{equation}
\begin{equation}
	\begin{aligned}
    M[D^{},B_c,\bar D^{*}]&=\int \frac{d^4k}{(2\pi)^4}i\cdot g_{D^{*}DP_8}\,g_{ D^* B_c P_{c}}^{1/2^{-}}\,p_{3\alpha} \,\bar{u}(p_4)\gamma^\nu \gamma_5(g_{\mu \nu}-\frac{p_{4\mu}p_{4\nu}}{m_{P_c}^2})(-g^{\alpha \mu}+\frac{k^\alpha k^\mu}{m_k^2}) (\not p_2 +m_2)\\&(A+B\gamma_{5})u(p_i)\frac{i^3 \mathcal{F}}{(p_1^2-m_1^2)(p_2^2-m_2^2)(k^2-m_k^2)}\,,
    \end{aligned}
\end{equation}
\begin{equation}
	\begin{aligned}
    M[D,B_c,B_c]
     &=\int \frac{d^4k}{(2\pi)^4}g_{P_8B_6B_{\bar 3}}\,g_{  P_{c}B_cD}^{1/2^{-}}\bar{u}(p_4)(\not k+m_k)\gamma_5(\not p_2+m_2)(A+B\gamma_{5})u(p_i)\\
     &\frac{i^3 \mathcal{F}}{(p_1^2-m_1^2)(p_2^2-m_2^2)(k^2-m_k^2)}\,,
    \end{aligned}
\end{equation}
\begin{equation}
	\begin{aligned}
    M[D^{*},B_c,B_c] &=\int \frac{d^4k}{(2\pi)^4}i\,g_{  P_{c} B_c D^*}^{1/2^{-}}\,g_{B_{\bar 3}B_6 P_8}\bar{u}(p_4)\gamma^\nu \gamma_5(g_{\mu \nu}-\frac{p_{4\mu}p_{4\nu}}{m_{P_c}^2})(\not k+m_k)\gamma_{5}(\not p_2+m_2)(-g^{\mu \alpha}+\frac{p_1^\alpha p_1^\mu}{m_1^2})\\
  &(A_1\gamma_\alpha\gamma_5+A_{2} \frac{p_{2\alpha}}{m_{i}} \gamma_{5}+B_{1} \gamma_{\alpha}+B_{2} \frac{p_{2\alpha}} {m_{i}})u(p_i)\frac{i^3 \mathcal{F}}{(p_1^2-m_1^2)(p_2^2-m_2^2)(k^2-m_k^2)}\,,
    \end{aligned}
\end{equation}
\begin{equation}
	\begin{aligned}
    M[P_8,B_8,V]&=\int \frac{d^4k}{(2\pi)^4}(-i)g_{P_cB_8V}^{1/2}\, g_{V P_8 P_8}\,\bar{u}(p_4)\gamma_5(g_{\mu\nu}-\frac{p_{4\mu} {p_4\nu}}{m_{P_c}^2})\gamma^\nu(\not p_2+m_2 )(-g^{\mu\alpha}+\frac{k^\mu k^\alpha}{m_k^2})
    \\&(A+B\gamma_{5})u(p_i)\left(p_{1\alpha}+p_{3\alpha}\right)\frac{i^3 \mathcal{F}}{(p_1^2-m_1^2)(p_2^2-m_2^2)(k^2-m_k^2)}\,.
    \end{aligned}
\end{equation}
\\
The amplitudes for $\Lambda^{0}_{b} \rightarrow P_c^{3/2^-}h^- $ are
\begin{equation}
	\begin{aligned}
 M[V,B_8,P_8]&=\int\frac{d^4k}{(2\pi)^4}(-1)\frac{g_{ P_8 B_8P_{c}}^{3/2^{-}}}{m_{P_8}^2} \,g_{VP_8P_8}\,\bar{u}_\alpha(p_4)\gamma_5 \not k \,k^\alpha(\not p_2+m_2)(-g^{\mu\nu}+\frac{p_1^\mu p_1^\nu}{m_1^2})\\
 &~~~~~(A_1\gamma_\mu\gamma_5+A_{2} \frac{p_{2\mu}}{m_{i}} \gamma_{5}+B_{1} \gamma_{\mu}+B_{2} \frac{p_{2\mu}} {m_{i}})u(p_i)(p_{3\nu}-k_\nu)\frac{i^3 \mathcal{F}}{(p_1^2-m_1^2)(p_2^2-m_2^2)(k^2-m_k^2)}\,,
    \end{aligned}
\end{equation}
\begin{equation}
	\begin{aligned}
M[D^{*},B_c,\bar D^{*}]&=\int\frac{d^4k}{(2\pi)^4}(-1)g_{ P_{c} B_c D^*  }^{3/2^{-}} \,g_{D^{*}D^*P_8}\,\bar{u}(p_4)(\not p_2+m_2)(-g_{\mu \beta}+\frac{k_\mu k_\beta}{m_k^2})(-g_{nt}+\frac{p_{1n}p_{1t}}{m_1^2}) \varepsilon^{mn\alpha\beta}\,p_{1m}\,k_{\alpha}\\
   &(A_1\gamma_t\gamma_5+A_{2} \frac{p_{2t}}{m_{i}} \gamma_{5}+B_{1} \gamma_{t}+B_{2} \frac{p_{2t}} {m_{i}})u(p_i)\frac{i^3 \mathcal{F}}{(p_1^2-m_1^2)(p_2^2-m_2^2)(k^2-m_k^2)}\,,
    \end{aligned}
\end{equation}
\begin{equation}
	\begin{aligned}
M[D^{},B_c,\bar D^{*}]
    &=\int \frac{d^4k}{(2\pi)^4}(-i) g_{D^{*}DP_8}\,g_{ P_{c} B_c D^* }^{3/2^{-}}\,p_{3\alpha} \,\bar{u}_\mu(p_4)(-g^{\alpha \mu}+\frac{k^\alpha k^\mu}{m_k^2}) (\not p_2 +m_2)(A+B\gamma_{5})\\
    &u(p_i)\frac{i^3 \mathcal{F}}{(p_1^2-m_1^2)(p_2^2-m_2^2)(k^2-m_k^2)}\,,
    \end{aligned}
\end{equation}
\begin{equation}
	\begin{aligned}
M[D^{*},B_c,B_c]&=\int \frac{d^4k}{(2\pi)^4} (-i)g_{  P_{c} B_c D^*}^{3/2^{-}}\,g_{B_{\bar 3}B_6 P_8}\bar{u}^\alpha(p_4) (\not k+m_k)\gamma_{5}(\not p_2+m_2)(-g_{\alpha \mu}+\frac{p_{1\alpha}p_{1\mu}}{m_1^2})\\
&(A_1\gamma_\mu\gamma_5+A_{2} \frac{p_{2\mu}}{m_{i}} \gamma_{5}+B_{1} \gamma_{\mu}+B_{2} \frac{p_{2\mu}} {m_{i}})u(p_i)\frac{i^3 \mathcal{F}}{(p_1^2-m_1^2)(p_2^2-m_2^2)(k^2-m_k^2)}\,,
    \end{aligned}
\end{equation}
\begin{equation}
	\begin{aligned}
M[P_8,B_8,V]&=\int \frac{d^4k}{(2\pi)^4}(-i)g^{3/2}_{P_cB_8V}\, g_{V P _8P_8}\,\bar{u}^\mu(p_4)(\not p_2 +m_2)(-g^{\mu \alpha}+\frac{k^\mu k^\alpha}{m_k^2})(A+B\gamma_{5})u(p_i) \\
&\left(p_{1\alpha}+p_{3\alpha}\right)\frac{i^3 \mathcal{F}}{(p_1^2-m_1^2)(p_2^2-m_2^2)(k^2-m_k^2)}\,,
    \end{aligned}
\end{equation}
\begin{equation}
	\begin{aligned}
    M[D^{*},B_c,\bar D]&=\int\frac{d^4k}{(2\pi)^4}\frac{g_{ P_{c} B_c D}^{3/2^{-}}}{m_D^2}\, g_{D^* D P_8}\,\bar u_\nu(p_4)\gamma_5\not k \,k^\nu (\not p_2+m_2)\,p_3^\alpha (-g_{\mu \alpha}+\frac{p_{1\mu}p_{1\alpha}}{m_1^2})\\
    &(A_1\gamma_\mu\gamma_5+A_{2} \frac{p_{2\mu}}{m_{i}} \gamma_{5}+B_{1} \gamma_{\mu}+B_{2} \frac{p_{2\mu}} {m_{i}})u(p_i)\frac{i^3 \mathcal{F}}{(p_1^2-m_1^2)(p_2^2-m_2^2)(k^2-m_k^2)}\,,
    \end{aligned}
\end{equation}
\begin{equation}
	\begin{aligned}
    M[D^{},B_c,B_c]&=\int \frac{d^4k}{(2\pi)^4}(-1)\frac{g_{ P_{c} B_c D}^{3/2^{-}}}{m_D^2}\,g_{P_8B_6B_{\bar 3}}\,\bar{u}_\nu(p_4)\gamma_5 \not p_1 \, p_1^\nu (\not k+m_k)\gamma_5(\not p_2+m_2)\\
    &(A+B\gamma_{5})u(p_i)\frac{i^3 \mathcal{F}}{(p_1^2-m_1^2)(p_2^2-m_2^2)(k^2-m_k^2)}\,.
    \end{aligned}
\end{equation}

\section{Full expressions of amplitudes}\label{app.C}
Here, we give the full amplitudes of all $\Lambda^{0}_{b}$ decay channels considered in this work:
\begin{equation}
 \begin{aligned}
 \mathcal{A}(\Lambda_b^0\to P_c^{1/2^-}\pi^-)&=\mathcal{M}(D^{*-},\Lambda_c^+,\bar{D}^0;P_c^{1/2^-})+\mathcal{M}(D^{*-},\Lambda_c^+,\bar{D}^{*0};P_c^{1/2^-})+\mathcal{M}(D^{-},\Lambda_c^+,\bar{D}^{*0};P_c^{1/2^-})\\
 &+\mathcal{M}(D^{-},\Lambda_c^+,\Sigma_c^{++};P_c^{1/2^-})
 +\mathcal{M}(D^{*-},\Lambda_c^+,\Sigma_c^{++};P_c^{1/2^-})+
 \mathcal{M}(\rho^-,p,\pi^0;P_c^{1/2^-})\\
 &+\mathcal{M}(\pi^-,p,\rho^0;P_c^{1/2^-})+\mathcal{M}(\rho^-,p,\omega;P_c^{1/2^-})\,,
    \end{aligned}
\end{equation}
\begin{equation}
 \begin{aligned}
 \mathcal{A}(\Lambda_b^0\to P_c^{3/2^-}\pi^-)&=\mathcal{M}(D^{*-},\Lambda_c^+,\bar{D}^0;P_c^{3/2^-})+\mathcal{M}(D^{*-},\Lambda_c^+,\bar{D}^{*0};P_c^{3/2^-})+\mathcal{M}(D^{-},\Lambda_c^+,\bar{D}^{*0};P_c^{3/2^-})\\
 &+\mathcal{M}(D^{-},\Lambda_c^+,\Sigma_c^{++};P_c^{3/2^-})
 +\mathcal{M}(D^{*-},\Lambda_c^+,\Sigma_c^{++};P_c^{3/2^-})+
 \mathcal{M}(\rho^-,p,\pi^0;P_c^{3/2^-})\\
 &+\mathcal{M}(\pi^-,p,\rho^0;P_c^{3/2^-})+\mathcal{M}(\rho^-,p,\omega;P_c^{3/2^-})\,,
    \end{aligned}
\end{equation}
\begin{equation}
 \begin{aligned}
 \mathcal{A}(\Lambda_b^0\to P_c^{1/2^-}K^-)&=\mathcal{M}(D_s^-,\Lambda_c^+,\bar{D}^{*0};P_c^{1/2^-})+\mathcal{M}(D_s^{*-},\Lambda_c^+,\bar{D}^{0};P_c^{1/2^-})+\mathcal{M}(D_s^{*-},\Lambda_c^+,\bar{D}^{*0};P_c^{1/2^-})\\
 &+\mathcal{M}(K^{*-},p,\pi^0;P_c^{1/2^-})+
 \mathcal{M}(K^{*-},p,\rho^0;P_c^{1/2^-})+
  \mathcal{M}(K^{*-},p,\omega;P_c^{1/2^-})\\
  &+\mathcal{M}(K^{-},p,\rho^0;P_c^{1/2^-})+
  \mathcal{M}(K^{-},p,\omega;P_c^{1/2^-})\,,
     \end{aligned}
\end{equation}
\begin{equation}
 \begin{aligned}
 \mathcal{A}(\Lambda_b^0\to P_c^{3/2^-}K^-)&=\mathcal{M}(D_s^-,\Lambda_c^+,\bar{D}^{*0};P_c^{3/2^-})+\mathcal{M}(D_s^{*-},\Lambda_c^+,\bar{D}^{0};P_c^{3/2^-})+\mathcal{M}(D_s^{*-},\Lambda_c^+,\bar{D}^{*0};P_c^{3/2^-})\\
 &+\mathcal{M}(K^{*-},p,\pi^0;P_c^{3/2^-})+
 \mathcal{M}(K^{*-},p,\rho^0;P_c^{3/2^-})+
  \mathcal{M}(K^{*-},p,\omega;P_c^{3/2^-})\\
  &+\mathcal{M}(K^{-},p,\rho^0;P_c^{3/2^-})+
  \mathcal{M}(K^{-},p,\omega;P_c^{3/2^-})\,.
     \end{aligned}
\end{equation}
\clearpage
\bibliographystyle{unsrturl}
\bibliography{References} 
\end{document}